# Non-coplanar spin structure in a metallic thin film of triangular lattice antiferromagnet CrSe

Yusuke Tajima[1], Junichi Shiogai[1,2,a)], Kohei Ueda[1,2], Hirotake Suzaki[1],

Kensuke Takaki[1], Takeshi Seki[3,4], Kazutaka Kudo[1,2], and Jobu Matsuno[1,2]

[1]*Department of Physics, Osaka University, Toyonaka, Osaka 560-0043, Japan*

[2]*Division of Spintronics Research Network, Institute for Open and Transdisciplinary Research Initiatives, Osaka University, Suita, Osaka 565-0871, Japan*

[3]*Institute for Materials Research, Tohoku University, Sendai 980-8577, Japan*

[4]*Center for Science and Innovation in Spintronics, Tohoku University, Sendai 980-8577, Japan*

[a)] Author to whom correspondence should be addressed.

Electronic mail: junichi.shiogai.sci@osaka-u.ac.jp

**ABSTRACT**

An antiferromagnetic metal with two-dimensional triangular network offers a unique playground of intriguing magneto-transport properties and functionalities stemming from interplay between conducting electrons and intricate magnetic phases. A NiAs-type CrSe is one of the candidates owing to alternate stackings of Cr and Se triangular atomic networks in its crystal structure. While fabrication of CrSe thin films is indispensable to develop functional devices, studies on its thin-film properties have been limited to date due to the lack of metallic samples. Here, we report on realization of metallic conductivities of CrSe thin films, which allows to investigate their intrinsic magneto-transport properties. The metallic sample exhibits co-occurrence of a weak ferromagnetism with perpendicular magnetic anisotropy and the antiferromagnetic behavior, indicating the presence of non-coplanar spin structures. In addition, control of polarity and tilting angle of the non-coplanar spin structure is accomplished by a sign of cooling magnetic fields. The observed non-coplanar spin structure, which can be a source of emergent magnetic field acting on the conducting electrons, highlights a high potential of the triangular lattice antiferromagnet and provide unique platform for functional thin-film devices composed of NiAs-type derivative Cr chalcogenides and pnictides.

**Main text**

## I.    Introduction

Triangular lattice antiferromagnets have long received attentions owing to various magnetic states [1,2] and related magneto-transport phenomena [3,4]. A key feature of the triangular lattice antiferromagnets is suppression of conventional long-range orders by geometrical frustration [5,6], which makes system be sensitive to magnetic anisotropy, thermal and quantum fluctuations, external magnetic field, and the other competing exchange interactions. When the frustrated spin system is subjected to these external stimuli, a rich variety of magnetic phases emerge such as non-colinear [7] and non-coplanar [8] spin structures, magnetic skyrmion [9,10], a chiral spin liquid state [11,12], and disordered states known as spin glass state [13,14]. Another intriguing feature, especially in non-coplanar magnets, is the emergent magnetic field associated with the scalar spin chirality [3], which corresponds to a solid angle subtended by neighboring spins so called the real-space Berry phase [15,16]. In some non-coplanar magnetic conductors, the connection of conducting electrons to the real-space emergent magnetic field is revealed as topological Hall effect [10,17,18,19,20], non-reciprocal magneto-resistance [21], and electromagnetic induction [22]. It has been suggested that properties of those exotic spin textures can be enriched in nanostructures [23,24,25] and heterointerfaces with conventional ferromagnets [26] or superconductors [27]. In this aspect, exploration of magneto-transport properties in thin films composed of triangular lattice antiferromagnets is a central topic for further functionalization of their exotic magnetic properties and magnetoelectric effects.

The NiAs-type and its derivative compounds Cr$X$ ($X$ = S, Se, Te, and Sb) exhibit diverse magnetism depending on the choice of chalcogen and pnictogen elements [28]. For instance, the CrTe shows a ferromagnetism in bulk and thin-film forms where ferromagnetic transition temperature and magnetic anisotropy can be tuned by the Cr and Te composition ratio [29,30,31]. The CrSb has received recent attentions as a new class of antiferromagnet, namely an altermagnet, in which the spin-split electronic band structure originates from not relativistic spin-orbit interaction but its crystal symmetry [32,33]. Thanks to the large spin-splitting band, the altermagnet can be a candidate material as an efficient spin polarizer [34]. Among various types of magnetic materials, we focus on an antiferromagnetic CrSe, which has been long known as a triangular lattice antiferromagnet [35]. Figure 1(a) illustrates crystal structure of the NiAs-type CrSe ($P6_3/mmc$) and suggested Cr spin configuration. The Cr and Se two-dimensional triangular networks are alternately stacked along the $c$ axis while its spin structure has not been fully understood. The first neutron scattering experiment for CrSe reported that Cr spins in a basal plane are canted out toward the same direction [35], forming so called umbrella-like spin structure as shown in Fig. 1(a). Later experiment performed for the NiAs-type derivative $Cr_2Se_3$ reported more complex non-coplanar spin structure [36]. In the Cr-Se binary phase diagram, there are some equilibrium intermetallic phases: $Cr_5Se_8$, $Cr_2Se_3$, $Cr_3Se_4$, and $Cr_7Se_8$, all of which are in the NiAs-type derivative crystal structure with ordered arrangement of Cr vacant sites [37], and a high-temperature NiAs-type ($P6_3/mmc$) and its derivative ($P\bar{3}m1$) CrSe phases [38]. All binary phases at low and high temperature have the identical framework of crystal structure, where Cr atoms form edge-sharing and face-sharing $CrSe_6$ octahedra in the $ab$ plane and along $c$ axis, respectively. The only differences among NiAs-type and its derivative Cr-Se compounds are fraction and position of Cr deficient sites. The Néel temperature increases from 43, 82, and 280 K for $Cr_2Se_3$, $Cr_3Se_4$, and CrSe, respectively, in the order of the filling fraction of Cr sites [39]. As for the electrical transport properties, while a metallic conduction has been reported in bulk single crystals [39,40,41,42], being consistent with semimetallic band structure in first-principles calculations [43,44], some experimental studies have reported semiconducting behavior in bulk [45,46] and thin-film samples [47,48,49,50] with a small band gap (a few tens meV) [46,47]. This is probably because the Fermi level and/or band gap are sensitive to a subtle off-stoichiometry or thickness in such a semimetal [43,44].

In particular, thin-film samples reported so far have exhibited only semiconducting behavior [47,48,49,50], hindering explicit determination of spin structure through the magneto-transport measurements. Nevertheless, some spin-related electrical transport experiments have been performed up to date through the interfacial effects showing exchange bias [48], topological Hall effect [49], and magnetic proximity effect [50] between an insulating CrSe layer and conducting ferromagnetic or topological materials. Within the context of non-coplanar spin structure having a scalar spin chirality as mentioned above, preparation of the metallic CrSe thin films and exploration of magneto-transport properties should be tackled for in-depth understanding of its spin structures.

In this Article, we have found evolution toward metallic conduction in CrSe thin films by increasing growth temperature. Metallic conduction allows to investigate coupling of electric transport properties with spin structures via their magneto-transport measurements. From magnetic-field and temperature dependences of longitudinal and Hall resistivities, coexistence of a weak ferromagnetism with perpendicular magnetic anisotropy and antiferromagnetic spin-flop transition accompanied with memory effect was found in the metallic thin-film sample. We concluded that these features originate from a non-coplanar spin structure of CrSe.

## II. Experimental methods

The CrSe thin films were grown on $Al_2O_3(0001)$ substrates by pulsed-laser deposition (PLD) at the substrate temperature $T_{sub} = 250, 350, 450, 550, 650, 750$, and $850^{\circ}$C. The CrSe and Se targets used in PLD were synthesized by pelletizing commercially available stoichiometric CrSe and Se powders (Kojundo Chemical Laboratory Co., Ltd), respectively. For the electrical transport measurement, 50-nm-thick CrSe thin films were prepared followed by deposition of 10-nm-thick amorphous Se cap layer at room temperature. For preparing Se-rich sample as a reference, the Se target was ablated when the sample was cooled to room temperature after the CrSe thin film was grown at $T_{sub} = 650^{\circ}$C or $750^{\circ}$C. The chemical composition ratio Se/Cr of the thin-film samples was determined by energy dispersive x-ray spectroscopy (EDX) and inductively coupled plasma optical emission spectroscopy (ICP-OES). For the ICP analysis, the 100-nm-thick samples without the cap layer were prepared at the identical growth condition. The crystallinity was evaluated by x-ray diffraction and atomic force micrography and thickness was determined by x-ray reflectivity.

Temperature and magnetic field dependences of longitudinal and Hall resistivities were measured by standard five-terminal measurement in $^4$He cryostat equipped with a superconducting magnet (Oxford Instruments, plc.). For subtracting the component of ordinary Hall effect from $\rho_{yx}(H)$, linear fit was applied for high-field region in $\mu_0 H > 5$ T for CrSe thin films grown at $T_{sub}$ = 350, 450, 550, and 650ºC and $\mu_0 H > 11$ T for $T_{sub}$ = 750ºC. Magnetization measurement was performed by MPMS3 (Quantum Design, Inc.) with superconducting quantum interference device.

## III.    Results and Discussion

### 1.  Thin Film Growth and Structural Characterization.

Figure 1(b) shows 2theta-omega scan of x-ray diffraction (XRD) pattern of the Se-capped CrSe thin film for $T_{sub}$ = 750ºC. Diffraction peaks from (00$n$) were clearly observed while no other diffraction peaks from the film were detected, indicating single-phase and $c$-axis-orientated growth with negligible secondary phase. Figure 1(c) shows the CrSe(002) diffraction peaks for $T_{sub}$ from 350 to 750ºC. Systematic dependence of the $c$-axis length on $T_{sub}$ was not apparently observed. Note that any clear crystalline XRD patterns were not detected for $T_{sub}$ = 250 and 850ºC. The CrSe(002) diffraction peak for $T_{sub}$ = 650ºC with cooling in Se shifts (denoted as "cooling in Se") toward a higher angle, indicating shrinkage of the $c$-axis length. Figure 1(d) and 1(e) summarizes $T_{sub}$ dependence of the Se/Cr composition ratio and $c$-axis length, respectively. With increasing $T_{sub}$, the Se/Cr composition ratio monotonically decreases as expected from the high vapor pressure of Se and approaches the stoichiometric composition of CrSe. This trend is qualitatively consistent with the Cr-Se phase diagram [51], where the Se/Cr composition approaches unity towards high temperature. The $c$-axis lengths for all the CrSe thin films are comparable to the bulk value of NiAs-type CrSe ($P6_3/mmc$, $c = 6.030$Å [Ref. 37]). In contrast, the $c$-axis length of CrSe thin film for $T_{sub} = 650$ and 750°C with cooling in Se was determined to be 5.764 and 5.752 Å, respectively and their Se/Cr ratio was about 1.6, which are comparable to the value of $Cr_2Se_3$ bulk single crystal ($I2/m$, $c = 5.764$ Å [Ref. 37]). Although further microscopic characterization and valence determination are necessary, it can be speculated that cooling in the Se-rich condition leads to the $Cr_2Se_3$ phase while thin-film growth at $T_{sub} = 750$°C without excess Se causes stabilization of the high-temperature CrSe phase according to the analysis of the chemical composition and lattice constant.

## 2. Variation of electrical transport properties of CrSe thin films on growth temperature.

Figure 2 shows evolution of electrical transport properties of CrSe thin films upon increase of $T_{sub}$. Figure 2(a) shows $T_{sub}$ dependences of Hall coefficient $R_H$ (left axis) and longitudinal resistivity $\rho_{xx}$ (right) measured at $T = 200$ K. Here, the Hall coefficient, *i.e.*, the inverse value of carrier density ($n$ or $p = 1/eR_H$ with $e$ being elementary charge), was extracted from linear fits to the magnetic-field dependence of the Hall resistivity $\rho_{yx}(H)$ in the high-field range. In our definition, the negative and positive values of $R_H$ correspond to electron and hole-dominated transport, respectively. For the low temperature growth at $T_{sub} = 350$ and $450^{\circ}C$, the $R_H$ shows the negative value. With increasing $T_{sub}$ above $550^{\circ}C$, the $R_H$ changes its sign and does not show a significant variation on further increase of $T_{sub}$. Concomitantly, the value of $\rho_{xx}$ at $T = 200$ K monotonically decreases with increase $T_{sub}$. The thermal variations of $R_H$ for the CrSe films for various $T_{sub}$ are shown in Fig. 2(b). The negative value of $R_H(T)$ for $T_{sub} = 350$ and $450^{\circ}C$ shows a semiconducting dependence on $T$ and reaches $R_H = -2.9$ and $-2.1 \times 10^{-3}$ cm$^3$C$^{-1}$ at $T = 4.2$ K, which corresponds to the electron density of 2.2 and $3.0 \times 10^{21}$ cm$^{-3}$, respectively. In contrast, thermal variation of $R_H(T)$ for $T_{sub} = 550$, 650, and $750^{\circ}C$ is not significant, which is typical behavior of semimetals composed of multiple bands such as CrTe [Ref. 31]. The temperature dependence of longitudinal resistivities [$\rho_{xx}(T)$ curve] of the CrSe films for various $T_{sub}$ are shown in Fig. 2(c). The CrSe thin films for $T_{sub} = 350$ and $450^{\circ}C$ exhibit insulating or semiconducting behavior (d$\rho_{xx}$/d$T$ < 0) in whole temperature range, respectively, being consistent with the semiconducting $R_H(T)$ discussed in Fig. 2(b). Indeed, we can estimate an activation energy to be 16.8 meV from the $R_H(T)$ in the samples for $T_{sub} = 350^oC$, being comparable with a previously reported band gap value of 34 meV in thin-film sample [47]. Note that such insulating behavior of $\rho_{xx}(T)$ curve of the CrSe-like thin film grown at low temperature is distinctly different from that for the $Cr_2Se_3$-like thin film (cooling in Se): $\rho_{xx}(T)$ of the $Cr_2Se_3$-like film more sharply increases in low temperature. For the higher $T_{sub}$ samples, on the other hand, the $\rho_{xx}(T)$ exhibits metallic behaviors ($d\rho_{xx}/dT > 0$) down to the lowest temperature. Based on temperature variation of $R_H$ and $\rho_{xx}$, the series of CrSe thin-film samples can be distinguished into the low-temperature semiconducting phase for $T_{sub} = 350$ and $450^oC$ and high-temperature metallic phase for $T_{sub} = 550, 650,$ and $750^oC$. Despite the significant structural difference was not observed from the present XRD analysis, we clearly observed an electronic phase transition by variation of the growth temperature. Further microscopic analysis may illustrate the link between NiAs-derivative crystal structures and electronic structures in high-temperature and low-temperature phases. Hereafter, we focus on the magneto-transport properties for the metallic CrSe thin film grown at $750^oC$.

### 3. Magneto-transport properties of metallic CrSe thin film.

The impact of the spin structure on the electrical transport in CrSe was investigated by resistivity measurement of metallic sample under magnetic field. Figure 3(a) shows $\rho_{xx}(T)$ of the CrSe thin film for $T_{sub} = 750^\circ C$ under the out-of-plane magnetic field of $\mu_0 H = 0$ T [the same trace presented in Fig. 2(c)] and 14 T. Here, the magnetoresistance (MR) is defined as the difference of $\rho_{xx}(T)$ under $\mu_0 H$ given by $\Delta\rho_{xx}(T) = \rho_{xx}(T, \mu_0 H = 14$ T$) - \rho_{xx}(T, \mu_0 H = 0$ T$)$, With decreasing $T$, the negative MR ($\Delta\rho_{xx} < 0$) appears blow 180 K, which corresponds to the temperature where a kink was observed in $\rho_{xx}(T)$ at zero field (vertical dashed line). The coincidence of the kink feature, often assigned to the Néel temperature in metallic CrSe bulk crystals [39,42], and appearance of the negative MR at the same temperature suggests that the observed negative MR is spin-related phenomena. Figure 3(b) shows temperature dependence of magnetization [$M(T)$ curve] measured at $\mu_0 H = 1$ T after 5 T field cooling. Despite the broad transition, the magnetization develops around $T \sim 180$ K, which further supports the relation of the kink and negative MR features in $\rho_{xx}(T)$ and magnetic transition. In addition, the slope of $M(T)$ curve becomes steeper below $T \sim 100$ K, implying that the perpendicular magnetic anisotropy with ferromagnetism develops as will be discussed in Fig. 3(c) and 3(d). Nevertheless, the value of measured magnetic moment is as small as 0.051 $\mu_B$/Cr at the lowest temperature. In the situation that Cr spins are aligned antiferromagnetically in the *ab*-plane, such a very small magnetic moment should originate from a projection of the canted Cr magnetic moments along the *c* axis. Considering the low-spin state of $3d^4$ electron configuration of $Cr^{2+}$ in CrSe, the residual moments of 0.051 $\mu_B$/Cr along the *c* axis out of the spin magnetic moment of 2 $\mu_B$/Cr yield a titling angle of Cr spins from the basal plane to be $\sin^{-1}\left(\frac{0.051}{2.0}\right) \sim 1.5^o$.

Figure 3(c) shows out-of-plane magnetic-field dependence of MR ratio at $T = 4.2$, 20, 40, 100, and 200 K. Here, we define the MR ratio as $\Delta\rho_{xx}/\rho_{xx}^0$ with $\Delta\rho_{xx} = \rho_{xx}(H) - \rho_{xx}^0$ and $\rho_{xx}^0 = \rho_{xx}(0)$. At $T = 200$ K, above transition temperature of 180 K, the $\Delta\rho_{xx}/\rho_{xx}^0$ exhibited a negative and parabolic dependence. The negative MR becomes more significant at $T = 100$ K, and a butterfly-shaped hysteresis clearly emerges when decreasing $T$ below 40 K, signaling the development of a ferromagnetic order. It is worth noting that hysteresis in $\Delta\rho_{xx}/\rho_{xx}^0$ was observed in the very wide range of magnetic field up to 14 T at $T = 4.2$ K, which is extraordinarily larger than characteristic magnetic fields such as coercivity and saturation magnetic fields of the typical ferromagnetic compounds [30,31]. The non-saturating negative MR can be ascribed to suppression of spin scattering due to reorientation of antiferromagnetically-coupled spins [45,52]. Figure 3(d) shows magnetic-field dependence of the anomalous Hall resistivity $\rho_{yx}^A(H)$ obtained by subtracting the $H$-linear component from $\rho_{yx}(H)$, as a contribution from the ordinary

Hall effect. The consistent hysteretic behavior of $\rho_{xx}(H)$ and $\rho_{yx}^{A}(H)$ accompanied

with a small fraction of remanent $\rho_{yx}^{A}(0)$ was observed below $T = 100$ K, which

indicates the development of the spontaneous ferromagnetism along the $c$ axis and is

consistent with the steep increase of magnetization below that temperature as shown in

Fig. 3(b). The $\rho_{yx}^{A}(H)$ at $T = 4.2$ K exhibits the non-monotonic hump character around

5 T, which cannot be assigned to the anomalous Hall effect due to the magnetic-field

dependence of the net out-of-plane magnetic moment. In the framework of the real-space

Berry phase picture, the emergent magnetic field is developed by non-coplanar spin

structures, which results in the non-monotonic hump structure, so called the topological

Hall effect [17,20]. Indeed, the observation of the non-monotonic $\rho_{yx}^{A}(H)$ is consistent with the topological Hall effect, suggesting the presence of the non-coplanar spin structure in our CrSe thin film.

For interpretation of the intricate behaviors of $\rho_{xx}(H)$ and $\rho_{yx}^{A}(H)$, angular dependence of MR was measured. Figure 3(e) shows the MR ratios $\Delta\rho_{xx}/\rho_{xx}^{0}$ for the angle $\theta = 0$, 45, and 90°. Here, $\theta_{H}$ is defined as the angle from the normal direction to the sample plane and $H$ is fixed perpendicular to the electric current. Interestingly, for the in-plane sweep ($\theta_{H} = 90°$), the $\Delta\rho_{xx}/\rho_{xx}^{0}$ for upward and downward sweeps collapse in the low-field regime ($\mu_0 H < 3$ T) and exhibits a positive MR while the hysteretic behavior remains in the high-field regime. The drastic change in the low-field MR with respect to $\theta_H$ is more visible in polar plots of $\Delta$MR at $\mu_0 H = 1.5$ T and 7.0 T, as shown in Fig. 3(f). Here, $\Delta$MR is defined by difference between the values of $\Delta\rho_{xx}/\rho_{xx}^0$ for the upward and downward sweeps [green and blue arrows in Fig. 3(e)]. The low-field $\Delta$MR (at $\mu_0 H = 1.5$ T) exhibits the strongly uniaxial anisotropy. For the high-field MR (at $\mu_0 H = 7.0$ T), in contrast, the $\Delta$MR shows less anisotropic dependence. The distinct difference in low-field and high-field MR in $\theta_H$ dependence recalls the presence of two magnetic interactions with different magnetic anisotropy. By considering the remanent anomalous Hall resistivity in Fig. 3(d), the low-field MR comes from the weak but spontaneous ferromagnetism from the canted Cr spins towards *c*-axis direction, which are ferromagnetically coupled between the basal planes. In contrast, we consider the nearly isotropic high-field component of MR with its hysteresis closing at very high magnetic field to be the character of triangular antiferromagnets. When antiferromagnets are magnetized by a sufficiently strong magnetic field, antiferromagnetically coupled spins undergo flops accompanied with a sudden rotation of Néel vector, which renders distinct anomalies in magnetic and MR properties [53,54,55]. Such spin-flop transition usually occurs only when the field is applied along the well-defined spin axis in colinear antiferromagnets. In the case of triangular antiferromagnets with competing interplane interaction, however, it is suggested that spin-flop transition may occur when the field is applied both parallel and perpendicular to the antiferromagnetic plane [56,57]. The step-like and hysteretic behavior in negative MR implies a first-order nature of this magnetic transition. Another possibility is a spin-glass state, which emerges when the geometrical frustration is subjected to some amount of site disorders [58], leading to a broad and unsaturated MR [59]. More detailed spin structure may be clarified by magnetization and magneto-transport measurement using high-field facilities.

In the presence of competing interplane weak ferromagnetism and intraplane geometrical frustration, the canting angle at the low temperature should depend on its history [60,61], in particular the magnetic field applied when the system was frozen. Figure 4 shows the out-of-plane MR at 4.2 K starting from the "unmagnetized" state obtained by zero-field cooling (ZFC) [Fig. 4(a)] and "magnetized" states by +14 and −14 T field cooling (FC) [Fig. 4(b) and 4(c)], respectively. The $\rho_{xx}(H)$ obtained from ZFC shows the negative MR from the initial state up to 14 T and exhibits an almost symmetric shape with respect to zero magnetic field. In this situation, the magnetic field drives the spin-flop transition from the compensated antiferromagnetic state after ZFC into a weak ferromagnetic phase as schematically shown in the insets. Please note that a small asymmetric composition may come from superposition of $\rho_{yx}(H)$. In contrast, the $\rho_{xx}(H)$ obtained from the +14 T-FC consists of an asymmetric component in addition to the negative MR, as shown in Fig. 4(b). The initial value of $\rho_{xx}$ at +14 T shows a lower resistivity than that for the ZFC. After a cycling magnetic field between +14 T and −14 T, $\rho_{xx}$ at +14 T returns to the comparable value with that obtained in ZFC. While the negative MR is independent of polarity of the cooling field, we found that the sign of the asymmetric component is inverted by the polarity change of the field, as shown in Fig. 4(c). The memory effect in MR polarities under the perpendicular magnetic field implies that the magnetic field in FC more effectively acts on the system than in ZFC to force the magnetic moment along the field. The observed non-saturating negative magnetoresistance accompanied with the memory effect implies antiferromagnetic features as reported in various non-coplanar and non-colinear antiferromagnetic conductors [45,59,60].

By considering co-occurrence of the weak ferromagnetism with uniaxial anisotropy perpendicular to the basal planes and the antiferromagnetic features, we conclude that our metallic CrSe thin film possesses the non-coplanar spin structure, as one of the ground states in antiferromagnetically coupled Cr triangular networks. It turns out that polarity and canting angle of the Cr spins, which is a source of the real-space emergent magnetic field, can be tuned by cooling under the magnetic field.

## IV.    Summary and outlook

In summary, we have obtained a semiconductor-to-metal transition in CrSe thin films by increasing growth temperature. The metallic CrSe thin film exhibits the hysteretic negative magnetoresistance, remanent anomalous Hall resistivity, and the small Cr magnetic moment when the perpendicular magnetic field is applied, revealing the presence of the weak ferromagnetic order with perpendicular magnetic anisotropy. In addition, unsaturated and isotropic hysteretic magnetoresistance showing the memory effect is a reminiscent of the competition between the intraplane geometrical frustration and interplane coupling. The co-existence of the weak ferromagnetism along perpendicular to the basal plane and geometrical frustration within the plane indeed suggests the presence of the non-coplanar spin structure in the two-dimensional triangular lattice. Considering a rich variety of magnetic states exhibited in the NiAs-type and its derivative Cr chalcogenides and pnictides, our observation of the non-coplanar spin structure in the CrSe thin film offers a unique platform to tailor intricate magnetic interaction and spintronic functionalities in the NiAs-based thin-film and heterostructure devices.

**ACKNOWLEDGMENTS**

Authors thank Masaki Nakano for fruitful discussion. This work was partly performed under the GIMRT Program of the Institute for Materials Research, Tohoku University (Proposal No. 202212-CRKEQ-0007). The ICP-OES analysis was carried out by Shigeru Tamiya at Osaka University Core Facility Center as a part of MEXT Project for promoting public utilization of advanced research infrastructure (Program for supporting construction of core facilities) Grant Number JPMXS0441200023. This work was supported by JSPS KAKENHI Grant Number JP21K18889, JP23H01686, JP19H05823, and JP22H01182.

**REFERENCES**

1. A. V. Chubukov and D. I. Golosov, "Quantum theory of an antiferromagnet on a triangular lattice in a magnetic field," J. Phys.:Condens. Matter **3**, 69 (1991).

2. D. Yamamoto, T. Sakurai, R. Okuto, S. Okubo, H. Ohta, H. Tanaka, and Y. Uwatoko, "Continuous control of classical-quantum crossover by external high pressure in the coupled chain compound $CsCuCl_3$," Nature Commun. **12**, 4263 (2021).

3. N. Nagaosa and Y. Tokura, "Emergent electromagnetism in solids," Phys. Scr. **T146**, 014020 (2012).

4. S. Nakatsuji, N. Kiyohara, and T. Higo, "Large anomalous Hall effect in a non-collinear antiferromagnet at room temperature," Nature **527**, 212 (2015).

5. A. P. Ramirez, "Strongly Geometrically Frustrated Magnets," Annu. Rev. Mater. Sci. **24**, 453 (1994).

6. C. Nisoli, R. Moessner, and P. Schiffer, "Colloquium: Artificial spin ice: Designing and imaging magnetic frustration," Rev. Mod. Phys. **85**, 1473 (2013).

7. K. Kubo and T. Z. Momoi, "Ground state of a spin system with two- and four-spin exchange interactions on the triangular lattice," Z. Phys. B Con. Mat. **103**, 485 (1997).

8. R. Shindou and N. Nagaosa, "Orbital Ferromagnetism and Anomalous Hall Effect in Antiferromagnets on the Distorted fcc Lattice," Phys. Rev. Lett. **87**, 116801 (2001).

9. Y. Tokura and N. Kanazawa, "Magnetic Skyrmion Materials,". Chem. Rev. **121**, 2857 (2021).

10. T. Kurumaji, T. Nakajima, M. Hirschberger, A. Kikkawa, Y. Yamasaki, H. Sagayama, H. Nakao, Y. Taguchi, T. Arima, and Y. Tokura, "Skyrmion lattice with a giant topological Hall effect in a frustrated triangular-lattice magnet," Science **365**, 914 (2019).

11. V. Kalmeyer and R. B. Laughlin, "Equivalence of the resonating-valence-bond and fractional quantum Hall states," Phys. Rev. Lett. **59**, 2095 (1987).

12. Y. Zhou, K. Kanoda, and T. K. Ng, "Quantum spin liquid states," Rev. Mod. Phys. **89**, 025003 (2017).

13. K. Binder and A. P. Young, "Spin glasses: Experimental facts, theoretical concepts, and open questions," Rev. Mod. Phys. **58**, 801 (1986).

14. U. Kamber, A. Bergman, A. Eich, D. Iuşan, M. Steinbrecher, N. Hauptmann, L. Nordström, M. I. Katsnelson, D. Wegner, O. Eriksson, and A. A. Khajetoorians, "Self-induced spin glass state in elemental and crystalline neodymium," Science **368**, 6494 (2020).

15. J. Ye, Y. B. Kim, A. J. Millis, B. I. Shraiman, P. Majumdar, and Z. Tešanović, "Berry Phase Theory of the Anomalous Hall Effect: Application to Colossal Magnetoresistance Manganites," Phys. Rev. Lett. **83**, 3737 (1999).

16. K. Ohgushi, S. Murakami, and N. Nagaosa, "Spin anisotropy and quantum Hall effect in the kagomé lattice: Chiral spin state based on a ferromagnet," Phys. Rev. B **62**, R6065 (2000).

17. Y. Taguchi, Y. Oohara, H. Yoshizawa, N. Nagaosa, and Y. Tokura, "Spin Chirality, Berry Phase, and Anomalous Hall Effect in a Frustrated Ferromagnet," Science **291**, 2573 (2001).

18. J. Matsuno, N. Ogawa, K. Yasuda, F. Kagawa, W. Koshibae, N. Nagaosa, Y. Tokura, and M. Kawasaki, "Interface-driven topological Hall effect in $SrRuO_3$-$SrIrO_3$ bilayer," Sci. Adv. **2**, e1600304 (2016).

19. Y. Ohuchi, J. Matsuno, N. Ogawa, Y. Kozuka, M. Uchida, Y. Tokura, and M. Kawasaki, "Electric-field control of anomalous and topological Hall effects in oxide bilayer thin films," Nature Commun. **9**, 213 (2018).

20. A. Neubauer, C. Pfleiderer, B. Binz, A. Rosch, R. Ritz, P. G. Niklowitz, and P. Böni, "Topological Hall Effect in the A Phase of MnSi," Phys. Rev. Lett. **102**, 186602 (2009).

21. T. Yokouchi, N. Kanazawa, A. Kikkawa, D. Morikawa, K. Shibata, T. Arima, Y. Taguchi, F. Kagawa, and Y. Tokura, "Electrical magnetochiral effect induced by chiral spin fluctuations," Nature Commun. **8**, 866 (2017).

22. T. Schulz, R. Ritz, A. Bauer, M. Halder, M. Wagner, C. Franz, C. Pfleiderer, K. Everschor, M. Garst, and A. Rosch, "Emergent electrodynamics of skyrmions in a chiral magnet," Nature Phys. **8**, 301 (2012).

23. A. O. Leonov and M. Mostovoy, "Edge states and skyrmion dynamics in nanostripes of frustrated magnets," Nature Commun. **8**, 14394 (2017).

24. P. Bruno, V. K. Dugaev, and M. Taillefumier, "Topological Hall Effect and Berry Phase in Magnetic Nanostructures," Phys. Rev. Lett. **93**, 096806 (2004).

25. N. Kanazawa, M. Kubota, A. Tsukazaki, Y. Kozuka, K. S. Takahashi, M. Kawasaki, M. Ichikawa, F. Kagawa, and Y. Tokura, "Discretized topological Hall effect emerging from skyrmions in constricted geometry," Phys. Rev. B **91**, 041122 (2015).

26. F. J. Morvan, H. B. Luo, H. X. Yang, X. Zhang, Y. Zhou, G. P. Zhao, W. X. Xia, and J. P. Liu, "An achiral ferromagnetic/chiral antiferromagnetic bilayer system leading to controllable size and density of skyrmions," Sci. Rep. **9**, 2970 (2019).

27. X. G. Wen, F. Wilczek, and A. Zee, "Chiral spin states and superconductivity," Phys. Rev. B **39**, 11413 (1989).

28. K. Adachi, "Magnetic Anisotropy Energy in Nickel Arsenide Type Crystals," J. Phys. Soc. Jpn. **16**, 2187 (1961).

29. H. Ipser, K. L. Komarek, and K. O. Kleep, "Transition metal-chalcogen systems viii: The Cr-Te phase diagram," J. Less-Common Met. **92**, 265 (1983).

30. Y. Fujisawa, M. Pardo-Almanza, J. Garland, K. Yamagami, X. Zhu, X. Chen, K. Araki, T. Takeda, M. Kobayashi, Y. Takeda, C. H. Hsu, F. C. Chuang, R. Laskowski, K. H. Khoo, A. Soumyanarayanan, and Y. Okada, "Tailoring magnetism in self-intercalated $Cr_{1+\delta}Te_2$ epitaxial films," Phys. Rev. Materials, **4**, 114001 (2020).

31. Y. Wang, S. Kajihara, H. Matsuoka, B. K. Saika, K. Yamagami, Y. Takeda, H. Wadati, K. Ishizaka, Y. Iwasa, and M. Nakano, "Layer-Number-Independent Two-Dimensional Ferromagnetism in $Cr_3Te_4$," Nano Lett. **22**, 9964 (2022).

32. L. Šmejkal, J. Sinova, and T. Jungwirth, "Emerging Research Landscape of Altermagnetism," Phys. Rev. X **12**, 040501 (2022).

33. S. Reimers, L. Odenbreit, L. Smejkal, V. N. Strocov, P. Constantinou, A. B. Hellenes, R. J. Ubiergo, W. H. Campos, V. K. Bharadwaj, A. Chakraborty, T. Denneulin, W. Shi, R. E. Dunin-Borkowski, S. Das, M. Kläui, J. Sinova, and M. Jourdan, "Direct observation of altermagnetic band splitting in CrSb thin films," arXiv:2310.17280.

34. R. González-Hernández, L. Šmejkal, K. Výborný, Y. Yahagi, J. Sinova, T. Jungwirth, and J. Železný, "Efficient Electrical Spin Splitter Based on Nonrelativistic Collinear Antiferromagnetism," Phys. Rev. Lett. **126**, 127701 (2021).

35. L. M. Corliss, N. Elliott, J. M. Hastings, and R. L. Sass, "Magnetic Structure of Chromium Selenide," Phys. Rev. **122**, 1402 (1961).

36. Y. Adachi, M. Ohashi, T. Kaneko, M. Yuzuri, Y. Yamaguchi, S. Funahashi, and Y. Morii, "Magnetic Structure of Rhombohedral $Cr_2Se_3$," J. Phys. Soc. Jpn, **63**, 1548 (1994).

37. F. H. Wehmeier, E. T. Keve, and S. C. Abrahams, "Preparation, structure, and properties of some chromium selenides. Crystal growth with selenium vapor as a novel transport agent," Inorg. Chem. **9**, 2125 (1970).

38. I. Tsubokawa, "The Magnetic Properties of Single Crystals of Chromium Selenide," J. Phys. Soc. Jpn. **15**, 2243 (1960).

39. S. Ohta, Y. Narui, and M. Ohwatari, "Electrical Properties of $Cr_{1-\delta}Se$ ($0.07 \leq \delta \leq 0.33$)," J. Phys. Soc. Jpn. **65**, 3084 (1996).

40. Y. Adachi, K. Izaki, K. Koike, H. Morita, T. Kaneko, H. Kimura, and A. Inoue, "Electrical resistivity and magnetic property for $Cr_2Se_3$ and its Te-substitution system," J. Magn. Magn. Mater. **310**, 1849 (2007).

41. Y. B. Li, Y. Q. Zhang, W. F. Li, D. Li, J. Li, and Z. D. Zhang, "Spin-glass-like behavior and electrical transport properties of $Cr_7(Se_{1-x}Te_x)_8$," Phys. Rev. B **73**, 212403 (2006).

42. J. Wu, C. L. Zhang, J. M. Yan, L. Chen, L. Guo, T. M. Chen, G. Y. Gao, L. Fei, W. Zhao, Y. Chai, and R. K. Zheng, "Magnetotransport and magnetic properties of the layered noncollinear antiferromagnetic $Cr_2Se_3$ single crystals," J. Phys.: Condens. Matter **32**, 475801 (2020).

43. J. Dijkstra, C. F. van Bruggeni, C. Haas, and R. A. de Groot, "Band-structure calculations, and magnetic and transport properties of ferromagnetic chromium tellurides (CrTe, $Cr_3Te_4$, $Cr_2Te_3$)," J. Phys.: Condens. Matter **1**, 9163 (1989).

44. Y. Gebredingle, M. Joe, and C. Lee, "First-Principles Calculations of the Spin-Dependent Electronic Structure and Strain Tunability in 2D Non-van der Waals Chromium Chalcogenides $Cr_2X_3$ (X = S, Se, Te): Implications for Spintronics Applications," ACS Appl. Nano Mater. **5**, 8 (2022).

45. S. J. Zhang, J. M. Yan, F. Tang, J. Wu, W. Q. Dong, D. W. Zhang, F. S. Luo, L. Chen, Y. Fang, T. Zhang, Y. Chai, W. Zhao, X. Wang, and R. K. Zheng, "Colossal Magnetoresistance in Ti Lightly Doped $Cr_2Se_3$ Single Crystals with a Layered Structure," ACS Appl. Mater. Interfaces **13**, 58949 (2021).

46. J. Yan, X. Luo, F. C. Chen, Q. L. Pei, G. T. Lin, Y. Y. Han, L. Hu, P. Tong, W. H. Song, X. B. Zhu, and Y. P. Sun, "Anomalous Hall effect in two-dimensional non-collinear antiferromagnetic semiconductor $Cr_{0.68}Se$," Appl. Phys. Lett. **111**, 022401 (2017).

47. A. Roy, R. Dey, T. Pramanik, A. Rai, R. Schalip, S. Majumder, S. Guchhait, and S. K. Banerjee, "Structural and magnetic properties of molecular beam epitaxy grown chromium selenide thin films," Phys. Rev. Materials **4**, 025001 (2020).

48. C. Wang, B. Zhang, B. You, S. K. Lok, S. K. Chan, X. X. Zhang, G. K. L. Wong, and I. K. Sou, "Molecular-beam-epitaxy-grown CrSe/Fe bilayer on GaAs(100) substrate," J. Appl. Phys. **102**, 083901 (2007).

49. J. H. Jeon, H. R. Na, H. Kim, S. Lee, S. Song, J. Kim, S. Park, J. Kim, H. Noh, G. Kim, S. K. Jerng, and S. H. Chun, "Emergent Topological Hall Effect from Exchange Coupling in Ferromagnetic $Cr_2Te_3$/Noncoplanar Antiferromagnetic $Cr_2Se_3$ Bilayers," ACS Nano **16**, 8974 (2022).

50. C. Y. Yang, L. Pan, A. J. Grutter, H. Wang, X. Che, Q. L. He, Y. Wu, D. A. Gilbert, P. Shafer, E. Arenholz, H. Wu, G. Yin, P. Deng, J. A. Borchers, W. Ratcliff II, and K. L. Wang, "Termination switching of antiferromagnetic proximity effect in topological insulator," Sci. Adv. **6**, eaaz8463 (2020).

51. H. Okamoto, *Data Handbook: Phase Diagrams for Binary Alloys 2nd Ed.,* (ASM International, Materials Park, OH, USA 1990).

52. J. Y. Chan, S. M. Kauzlarich, P. Klavins, R. N. Shelton, and D. J. Webb, Colossal negative magnetoresistance in an antiferromagnet. Phys. Rev. B **57**, R8103(R) (1998).

53. Y. Deng, Y. Yu, M. Z. Shi, Z. Guo, Z. Xu, J. Wang, X. H. Chen, and Y. Zhang, "Quantum anomalous Hall effect in intrinsic magnetic topological insulator $MnBi_2Te_4$," Science **367**, 895 (2020).

54. J. Wang, J. Deng, X. Liang, G. Gao, T. Ying, S. Tian, H. Lei, Y. Song, X. Chen, J. Guo, and X. Chen, "Spin-flip-driven giant magnetotransport in A-type antiferromagnet $NaCrTe_2$," Phys. Rev. Materials **5**, L091401 (2021).

55. D. G. Oh, J. H. Kim, M. K. Kim, K. W. Jeong, H. J. Shin, J. M. Hong, J. S. Kim, K. Moon, N. Lee, and Y. J. Choi, "Spin-flip-driven anomalous Hall effect and anisotropic magnetoresistance in a layered Ising antiferromagnet," Sci. Rep. **13**, 3391 (2023).

56. T. Nikuni and H. Shiba, "Quantum Fluctuations and Magnetic Structures of $CsCuCl_3$ in High Magnetic Field," J. Phys. Soc. Jpn. **62**, 3268 (1993).

57. Z. Jin, Z. C. Xia, M. Wei, J. H. Yang, B. Chen, S. Huang, C. Shang, H. Wu, X. X. Zhang, J. W. Huang, and Z. W. Ouyang, "3D spin-flop transition in enhanced 2D layered structure single crystalline $TlCo_2Se_2$," J. Phys.: Condens. Matter **28**, 396002 (2016).

58. P. A. Beck, "Properties of mictomagnets (spinglasses)," Prog. Mater. Sci. **23**, 1 (1980).

59. C. L. Canedy, K. B. Ibsen, G. Xiao, J. Z. Sun, A. Gupta, and W. J. Gallagher, "Magnetotransport and hysteretic behavior in epitaxial $La_{0.67}Ca_{0.33}MnO_{3-\delta}$ films," J. Appl. Phys. **79**, 4546 (1996)

60. T. C. Fujita, Y. Kozuka, M. Uchida, A. Tsukazaki, T. Arima, and M. Kawasaki, "Odd-parity magnetoresistance in pyrochlore iridate thin films with broken time-reversal symmetry," Sci. Rep. **5**, 9711 (2015).

61. J. Nogués and I. K. Schuller, "Exchange bias," J. Magn. Magn. Mater. **192**, 203 (1999).

62. K. Momma and F. Izumi, "*VESTA* 3 for three-dimensional visualization of crystal, volumetric and morphology data," J. Appl. Crystallogr. **44**, 1272 (2011).

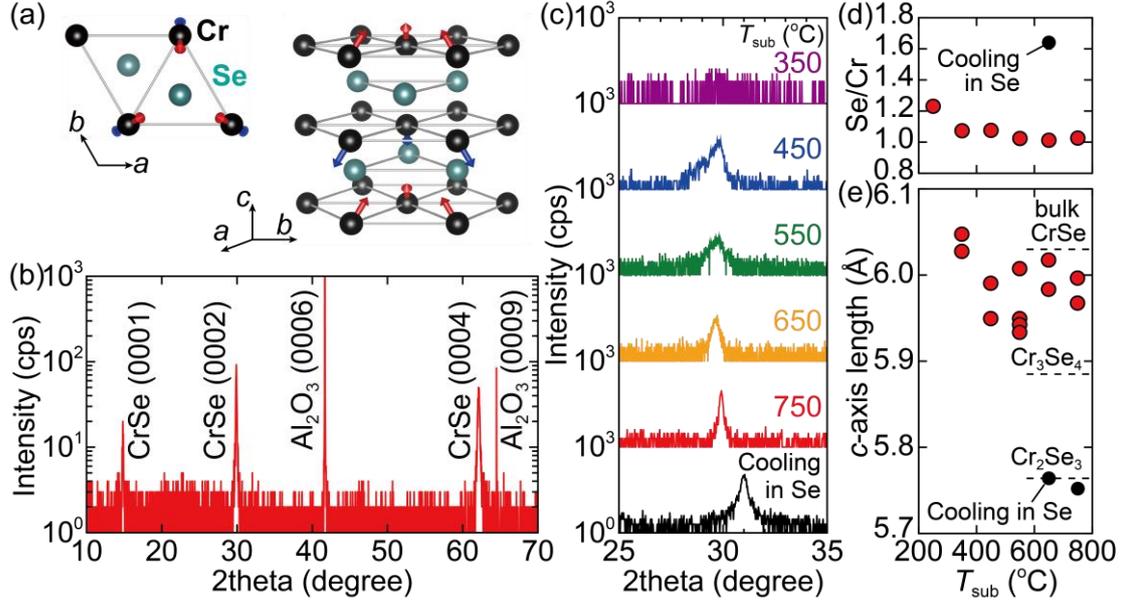

**FIG. 1.** Structural characterization of CrSe thin films. (a) Schematic crystal and spin structures of NiAs-type CrSe drawn by VESTA [62]. Arrows illustrate Cr spins, which form non-coplanar spin structure based on Ref. 35. (b) 2theta-omega scan of x-ray diffraction (XRD) pattern of Se-caped CrSe thin film grown at $T_{sub}$ = 750ºC. (c) 2theta-omega XRD patterns around CrSe(002) for the CrSe thin films grown at $T_{sub}$ = 350 (from top panel), 450, 550, 650, and 750ºC. As a reference, the XRD pattern of the Cr-Se thin film grown at $T_{sub}$ = 650ºC and cooled under the Se ablation is shown in the bottom panel. The $T_{sub}$ dependences of (d) Se/Cr chemical composition ratio and (e) $c$-axis length determined by CrSe(002) XRD peak. Black circles correspond to the data points obtained by cooling in Se. Horizontal dashed lines in (e) indicate the bulk values of CrSe, $Cr_3Se_4$, and $Cr_2Se_3$, respectively [37].

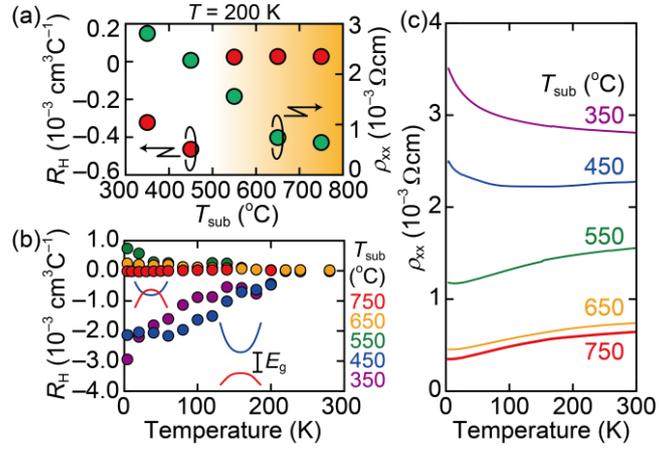

**FIG. 2.** Electrical transport characteristics. (a) Growth temperature $T_{sub}$ dependence of high-field Hall coefficient $R_H$ (red circles in left axis) and resistivity $\rho_{xx}$ (green circles in right axis) measured at $T = 200$ K. Temperature dependence of (b) $R_H$ and (c) $\rho_{xx}$ for $T_{sub} = 350$ (blue), 450 (green), 550 (yellow), 650 (orange), and 750°C (red).

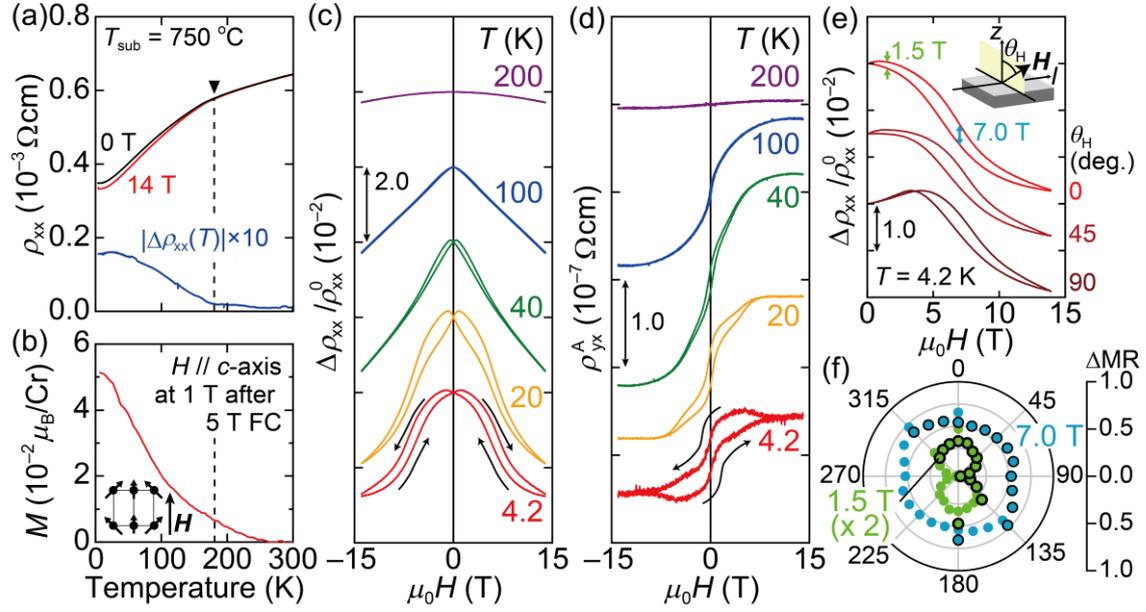

**FIG. 3.** Magneto-transport properties of the metallic CrSe thin film. (a) Temperature dependence of resistivity $\rho_{xx}$ under zero field (black) and 14 T magnetic field (red). Absolute difference between the values of $\rho_{xx}$ at 0 T and 14 T is plotted in blue solid line, where the vertical scale is multiplied by 10 for clarity. (b) Temperature dependence of magnetization in 1 T after 5 T field cooling (FC). (c) Magneto-resistance defined as $\Delta\rho_{xx}/\rho_{xx}^0$ and (d) magnetic-field dependence of anomalous Hall resistivity $\rho_{yx}^A$ at $T =$ 4.2 (red), 20 (orange), 40 (green), 100 (blue), and 200 K (purple). Here, $\Delta\rho_{xx}(H)$ and $\rho_{xx}^0$ represent $\rho_{xx}(H) - \rho_{xx}(0)$ and $\rho_{xx}(0)$. The $\Delta\rho_{xx}(H)/\rho_{xx}^0$ and $\rho_{yx}^A(H)$ curves are vertically offset for clarity. (e) $\Delta\rho_{xx}/\rho_{xx}^0$ $(H)$ for different magnetic-field angles $\theta_H =$ 0º, 45º and 90º with respect to the normal to the plane and being fixed perpendicular to the electric current at 4.2 K. (f) Polar plot of $\Delta$MR, which is defined as the difference in $\Delta\rho_{xx}/\rho_{xx}^0$ for upward and downward sweeps at 1.5 T (light green circles with black border) and 7.0 T (cyan circles with black border). The light green and cyan circles without black border are obtained by replotting the data for $\theta_H$ on $\theta_H+180^{\circ}$. Note that the radial scale of $\Delta\rho_{xx}/\rho_{xx}^0$ for 1.5 T is doubled for clarity.

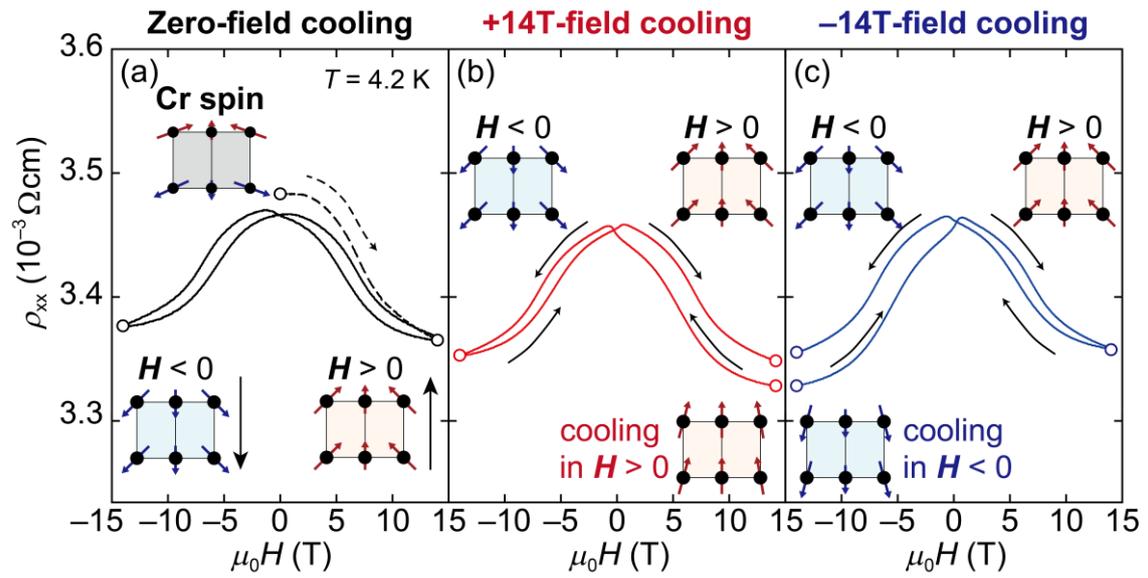

**FIG. 4.** Magnetic-field dependence of resistivity after (a) zero-field cooling, (b) field cooling at +14 T, and (c) field cooling at −14 T. Dashed line in (a) represents the initial curve obtained after the zero-field cooling. Inset shows schematics of proposed Cr spin structure at each magnetic field represented by the open circles (not to scale).